\documentclass{article}

\usepackage{PRIMEarxiv}

\usepackage[utf8]{inputenc} 
\usepackage[T1]{fontenc}    
\usepackage{hyperref}       
\usepackage{url}            
\usepackage{booktabs}       
\usepackage{amsfonts}       
\usepackage{nicefrac}       
\usepackage{microtype}      
\usepackage{lipsum}
\usepackage{fancyhdr}       
\usepackage{graphicx}       
\graphicspath{{media/}}     
\usepackage{cite}
\usepackage{amsmath,amssymb}
\usepackage{algorithmic}
\usepackage{subcaption}
\usepackage{textcomp}
\usepackage{xcolor, url}
\usepackage{multirow}
\usepackage{hyperref}

\title{Observations on LLMs for Telecom Domain: Capabilities and Limitations
\thanks{
Submitted for review to GLOBECOM 2023.
} 
}

\author{
  Sumit Soman \\
  Global AI Accelerator \\
  Ericsson \\
  Bangalore, India\\
  \texttt{sumit.soman@ericsson.com} \\
   \And
  Ranjani H G \\
  Global AI Accelerator \\
  Ericsson \\
  Bangalore, India\\
  \texttt{ranjani.h.g@ericsson.com} \\
}

\begin{document}
\maketitle

\begin{abstract}
The landscape for building conversational interfaces (chatbots) has witnessed a paradigm shift with recent developments in generative Artificial Intelligence (AI) based Large Language Models (LLMs), such as ChatGPT by OpenAI (GPT3.5 and GPT4), Google's Bard, Large Language Model Meta AI (LLaMA), among others. In this paper, we analyze capabilities and limitations of incorporating such models in conversational interfaces for the telecommunication domain, specifically for enterprise wireless products and services. Using Cradlepoint's publicly available data for our experiments, we present a comparative analysis of the responses from such models for multiple use-cases including domain adaptation for terminology and product taxonomy, context continuity, robustness to input perturbations and errors. We believe this evaluation would provide useful insights to data scientists engaged in building customized conversational interfaces for domain-specific requirements.
\end{abstract}


\section{Introduction \label{sec:intro}}
There has been significant traction in the development of Large Language Models (LLMs) recently, particularly generative Artificial Intelligence (AI) based LLMs. OpenAI introduced ChatGPT\footnote{\url{https://openai.com/blog/chatgpt}} \cite{stiennon2020learning, gao2022scaling}, and subsequently ChatGPT Plus\footnote{\url{https://openai.com/blog/chatgpt-plus}} based on GPT3.5 and GPT4 \cite{openai2023gpt4}. Google released Bard\footnote{\url{https://bard.google.com/}}, based on Language Model for Dialogue Applications (LaMDA) \cite{googlelamda}. Other notable efforts include LLaMA \cite{touvron2023llama}, Chinchilla \cite{hoffmann2022training}, PaLM \cite{chowdhery2022palm}, UL2 \cite{tay2022ul2}, Cerebras-GPT \cite{dey2023cerebras}. Literature on LLM based chat interfaces such as ChatGPT and Bard \cite{ram2023artificial, rahaman2023ai} discuss capabilities and prospects, such as for legal use-cases \cite{hoppner2023chatgpt}. With general availability of user interfaces, models and datasets \cite{kopf2023openassistant}, the development of (and scope for using) conversational interfaces has gained interest, as these models enable question answering for performing various tasks \cite{richards2023auto}. Motivated by recent evaluation reports \cite{bubeck2023sparks, martin2023hello, sejnowski2023large}, we investigate domain-specific capabilities and limitations of some models.

Our specific focus in this paper is on conversational assistants (or interfaces) for the telecom domain. A common use-case for organizations with enterprise (as well as consumer grade) products and associated services is to have a conversational assistant that can help users with tasks such as finding information about products and services, initial support for installation and configuration, operational use-cases like troubleshooting or performance monitoring, among others. In such cases, a conversational interface is the first line of interaction with the end-user. Thus, it becomes important for the chatbot (that often use LLMs) to understand domain terminology, concepts and context in such conversations. These aspects are considered in our work using a specific example for enterprise wireless products. However, in principle, the study and findings can be extended to other domains where training or fine-tuning domain-specific conversational models  would be of interest.

We aim to address the following Research Questions (RQ):
\begin{itemize}
    \item \textbf{RQ1:} Can conversational interfaces that use generative AI LLMs  adapt to questions related to domain specific terminology? E.g. telecom domain and product queries.
    \item \textbf{RQ2:} How do these models fare in retaining context(s) across conversations? E.g. co-reference resolution from queries and long-term context retention.
    \item \textbf{RQ3:} Are the recipes (responses provided as steps to be followed by the user) generated by such models accurate and reliable, or, do they tend to hallucinate?
    \item \textbf{RQ4:} Are these models robust to language or grammatical perturbations and can they adapt to specific domains?
\end{itemize}

To address these questions, we devise and conduct experiments using multiple generative AI models, including GPT4, GPT3.5, Bard (based on LaMDA) and LLaMA provided through HuggingChat\footnote{\url{https://huggingface.co/chat/}}. Our experiments are mapped to the respective RQs raised above. Having identified the typical requirements for wireless Enterprise chat interfaces, we wish to highlight that these requirements involve domain intensive terminology and (organization) specific products. The purpose of this paper is to share the observations on strengths and limitations of some LLMs as a user facing touch-point. Owing to the domain terminology and specific products, we think it may be injudicious to evaluate the models (using APIs) on large internal datasets without uncovering initial findings. The discussion in this paper is based on the responses received from the chat interfaces of the respective models. Training data used for different models and their token lengths are listed in Table \ref{tab:training_data}. We believe this captures the popular spectrum of generative AI LLMs available at this point of time (the experiments have been conducted during March - April 2023).  

\begin{table}[ht]

    \centering
    \scalebox{1.0}{
    \begin{tabular}{|c|c|c|c|}
    \hline
    \textbf{Model} & \textbf{Training data} & \textbf{Token Length} \\
    \hline
    GPT 3.5 (175 B)& Sept 2021 & 4k \\\hline
    GPT 4 (1 T) & Sept 2021 & 8k/32k \\\hline
    Bard/LaMDA (137 B) & 2022 (First Half)& 512\\\hline
    oasst-sft-6-llama-30b (65 B)& March 2023 &  4k\\\hline
    \end{tabular}
    }
    \caption{Training data cut-off \& token length of the LLMs}
    \label{tab:training_data}
\end{table}
\color{black}


\section{Experiments \label{sec:expts}}

We evaluate the models for multiple question-answering tasks, as detailed in the following sub-sections. The experiments are categorized into domain Question-and-Answer (Q\&A) (E1), product Q\&A (E2), context continuity (E3 - emulating context continuity for a product information query sequence, and E4 - emulating context continuity in a troubleshooting scenario) and robustness to spelling (or language) perturbations (E5).

\subsection{Domain Q\&A}
This pertains to question-answering task related to telecom domain (to address RQ1), including aspects such as network technology (4G, 5G etc.) and capabilities. The answers are evaluated based on the ability to discern such questions from common vocabulary. This is an important challenge when we use such LLMs for domain specific tasks. We note that though LLMs may have access to the domain (owing to the large and diverse training data), yet they may not be trained on domain-specific vocabulary, often due to dependence on Subject Matter Experts (SMEs) or lack of standardized domain taxonomy. We include four questions as shown in Table \ref{tab:ques_domain_qa} based on 5G technology\footnote{The deployment of 5G picked up pace from 2019. Most LLMs are trained using data upto 2021 (\ref{tab:training_data} and hence the models are expected to be able to provide answers.}. Questions 1-3 are related to general telecom domain, while Q4 is related to modem capability. Assessment of responses to these questions may be subjective in nature, as such we also evaluated Mean Opinion Score (MOS) and inter-rater agreement to quantitatively substantiate our findings.

\begin{table}[htbp]
\centering
\scalebox{1.0}{%
\begin{tabular}{|ll|}
\hline
\multicolumn{2}{|c|}{\textbf{E1: Domain Q\&A}}                                                                  \\ \hline
\multicolumn{1}{|l|}{Q1} & What are the different 5G spectrum layers?                                  \\ \hline
\multicolumn{1}{|l|}{Q2} & What are the different 5G architectures?                                    \\ \hline
\multicolumn{1}{|l|}{Q3} & What are the uses of mid-band 5G?                                           \\ \hline
\multicolumn{1}{|l|}{Q4} & Can we have a single modem, steering between LTE PDNs or 5G network slices?\\ \hline
\end{tabular}
}
\caption{Questions evaluated for domain Q\&A. \label{tab:ques_domain_qa}}
\end{table}

\subsection{Product Q\&A}

The other aspect related to domain adaptation is comprehending products (names as well as model names, components, specifications etc.) correctly. This is relevant since the questions from users may relate to detailed information about the products they use or intend to purchase, or pertain to operational aspects, such as installation, configuration, troubleshooting, among others. There may also be questions related to comparison of products, in which case the LLM should be able to interpret product names and specifications. We evaluate the seven questions shown in Table \ref{tab:ques_product_qa} for representative Cradlepoint products (whose datasheets are available publicly). One of the questions (Q5) pertains to a non-existent product model number at the time of evaluation. The purpose of placing this question in proximity to a similar question (Q4) with a correct product model number is to observe the ability of LLMs to discern factually incorrect questions, as well as possible confidence of the response henceforth. This experiment maps to RQ1 and RQ3 (for recipe generation - Q3 and hallucination - Q4 and Q5).

\begin{table}[hbtp]
\centering
\scalebox{1.0}{%
\begin{tabular}{|ll|}
\hline
\multicolumn{2}{|c|}{\textbf{E2: Product Q\&A}}                                                                  \\ \hline
\multicolumn{1}{|l|}{Q1} & How many expansion slots are there in E300 and what types of slots are they?                                  \\ \hline
\multicolumn{1}{|l|}{Q2} & What are the operating conditions for IBR900 router?                                    \\ \hline
\multicolumn{1}{|l|}{Q3} & What are the steps to setup IBR900?                                           \\ \hline
\multicolumn{1}{|l|}{Q4} & What is the power consumption of IBR1700 as per product specifications? \\ \hline
\multicolumn{1}{|l|}{Q5} & What is the power consumption of IBR700 as per product specifications? \\ \hline
\multicolumn{1}{|l|}{Q6} & Which are the ruggedized routers of Cradlepoint? \\ \hline
\multicolumn{1}{|l|}{Q7} & Compare the E300 and IBR900 router specifications? \\ \hline
\end{tabular}
}
\caption{Questions evaluated for product Q\&A \label{tab:ques_product_qa}}
\end{table}

\subsection{Context Continuity}

Context continuity is important for conversational interfaces for better user experience, as mentioned in RQ2. This refers to the ability to retain context in user conversation (queries) from previous queries. We evaluate two flavors for context continuity. Table \ref{tab:ques_context_1} (E3) relates to queries on product specifications where subsequent queries need to retain product information from Q1. In Q3 to Q5 of E3, an intentional input perturbation as typographical error for LTE bands (as \textit{``let''} bands) is also introduced in this evaluation sequence, which is relevant to RQ4. Q4 and Q5 of E3 pertains to the ability to discern differences between product and model. In Table \ref{tab:ques_context_2} (E4), we simulate a troubleshooting conversation where the user asks multiple queries related to issues with a router and the domain context needs to be retained across the questions. We expect most of the responses to be drawn from the publicly available quick start guides provided for the product. 

\begin{table}[hbtp]
\centering
\scalebox{1.0}{%
\begin{tabular}{|ll|}
\hline
\multicolumn{2}{|c|}{\textbf{E3: Context Continuity (Product Information)}}                                                                  \\ \hline
\multicolumn{1}{|l|}{Q1} & Can we monitor cellular health as a service for IBR700?                                  \\ \hline
\multicolumn{1}{|l|}{Q2} & Can we check using web links?                                    \\ \hline
\multicolumn{1}{|l|}{Q3} & Does this inform about let bands used? If yes, list the routers?                                           \\ \hline
\multicolumn{1}{|l|}{Q4} & Does this inform about the let bands used for 1200M modem? \\ \hline
\multicolumn{1}{|l|}{Q5} & Does this inform about the let bands used for 1200M-B modem? \\ \hline
\end{tabular}
}
\caption{Context Continuity - 1 \label{tab:ques_context_1}}
\end{table}

\begin{table}[hbtp]
\centering
\scalebox{1.0}{%
\begin{tabular}{|ll|}
\hline
\multicolumn{2}{|c|}{\textbf{E4: Context Continuity (Troubleshooting)}}                                                                  \\ \hline
\multicolumn{1}{|l|}{Q1} & \begin{tabular}{@{}c@{}}I have a Cradlepoint E300 router whose cellular health LED shows one\\blinking bar and power LED is yellow. What should I do? \end{tabular}                               \\ \hline
\multicolumn{1}{|l|}{Q2} & Which LED will show signal strength?                                    \\ \hline
\multicolumn{1}{|l|}{Q3} & How do I update its firmware?                                           \\ \hline
\multicolumn{1}{|l|}{Q4} & Which card can I insert and what should I check for? \\ \hline
\multicolumn{1}{|l|}{Q5} & Can you give me the web link of the document for troubleshooting it? \\ \hline
\multicolumn{1}{|l|}{Q6} & \begin{tabular}{@{}c@{}}Is it possible to factory reset? Can you point me to the \\ document URL for my router? \end{tabular}  \\ \hline
\multicolumn{1}{|l|}{Q7} & What problem was I facing earlier? \\ \hline
\end{tabular}
}
\caption{Context Continuity - 2 \label{tab:ques_context_2}}
\end{table}

\subsection{Input Perturbations (Language Errors)}

This experiment aims to address RQ4. The questions in Table \ref{tab:ques_lang_err} include common language errors, related to both general English language usage, as well as domain terminology - protocols like Internet Key Exchange (IKE) and Internet Protocol Security (IPSec). The errors, indicated in \textit{italics}, are also generated based on keyboard distance.
\begin{table}[hbtp]
\centering
\scalebox{1.0}{%
\begin{tabular}{|ll|}
\hline
\multicolumn{2}{|c|}{\textbf{E5: Language Errors}}                                                                  \\ \hline
\multicolumn{1}{|l|}{Q1} & \textit{Whact} is the vpn \textit{tunhel counvt} in W2005?                                  \\ \hline
\multicolumn{1}{|l|}{Q2} & Is \textit{locatoon servcies} included in my \textit{subcrption}?                                    \\ \hline
\multicolumn{1}{|l|}{Q3} & Can we monitor cellular \textit{hdalth} as a \textit{servkce} for IBR1700?                                           \\ \hline
\multicolumn{1}{|l|}{Q4} & Can we \textit{condifure} \textit{ispec}? \\ \hline
\multicolumn{1}{|l|}{Q5} & Can we \textit{condifure} ike? \\ \hline
\multicolumn{1}{|l|}{Q5} & Can we \textit{confifure} ike? \\ \hline
\multicolumn{1}{|l|}{Q5} & Can we \textit{cahnge} the \textit{aldrting tije peeiod}? \\ \hline
\end{tabular}
}
\caption{Language errors (spelling perturbations). \label{tab:ques_lang_err}}
\end{table}

\subsection{Parameters and Prompts Used}

We describe the model parameters set for the experiments reported in this work in Table \ref{tab:modelparams}. For fairness of comparison, we have chosen default parameter settings for all cases. The trial version of Bard has been used (there is no configuration option available in the web-based user interface). Similarly for LLaMA, the interface through HuggingChat provides \textit{oasst-sft-6-llama-30b} model and does not include any configurable parameter settings.

\begin{table}[h]
\centering
\begin{tabular}{|p{1.5in}|p{4.5in}|}
\hline
\multicolumn{1}{|c|}{\textbf{Model}} & \multicolumn{1}{c|}{\textbf{Parameters}}                                       \\ \hline
ChatGPT Plus (gpt-4)                               & Temperature = 0.5, Max Length = 2048,  Top P = 1, Frequency Penalty = 0, Presence Penalty = 0 \\ \hline
ChatGPT (gpt-3.5-turbo)                            & Temperature = 0.5, Max Length = 2048, Top P = 1, Frequency Penalty = 0, Presence Penalty = 0 \\ \hline
Bard                                               &                                                                                              \\ \cline{1-1}
HuggingChat        & \multirow{-2}{*}{User cannot configure parameters (as on April 2023).}                     \\ \hline
\end{tabular}
\caption{Model parameters used for experiments.\label{tab:modelparams}}
\end{table}
\color{black}

The prompts used are common across all the LLMs for a fair comparison. For E1, the following prompt is used - \textit{``You are an AI assistant for me. You are given a telecom related question. Provide a short answer. If you don't know the answer, just say ``I do not know.'' Do not try to make up an answer. If the question is not about telecom, politely inform them that you are tuned to only answer questions about telecom.''}.

For E2-E5, the prompt used is \textit{``You are an AI assistant for me. The documentation is located at https://customer.cradlepoint.com. You are given a question. Provide a conversational multi-step answer. You should only use content that is in the Cradlepoint URL. If you don't know the answer, just say ``I do not know.'' Do not try to make up an answer. If the question is not about Cradlepoint, politely inform them that you are tuned to only answer questions about Cradlepoint.''}

\color{blue}
\color{black}

\section{Results and Discussion \label{sec:results}}
We present our observations on the suitability of responses for each RQ based on the corresponding experiment. The responses are analysed for each LLM in this section.

\subsection{RQ1 - E1 and E2}
\label{ssec:rq1}
For E1, Q1, Bard identifies the three bands as low, mid and high, but does not provide specific details of frequency bands. Instead, it uses comparative adverbs to describe the pros and cons of each band,  thus making the answer descriptive. GPT3.5 simply names the bands, no details about frequency bands are provided, while GPT4's response includes both the categorization and frequency range for the respective bands, as well as a description of pros and cons of each band. It identifies the bands as below 1~GHz, 1-6~GHz and above 6 GHz. LLaMA, on the other hand, identifies the frequency bands as below 600 MHz, 600 MHz - 24 GHz and above 24 GHz, but does not provide any comparison of pros and cons. As the frequency band description may vary across multiple data sources, for a na\"ive user, the information about pros and cons of the respective bands would be useful. It is however, equally important, to provide accurate information of frequency bands.

For Question 2 (related to 5G architecture), Bard identifies Stand-Alone (SA) and Non Stand-Alone (NSA) as possible architectures and elucidates on definition and benefits. Another observation from Bard's response is that it includes an additional paragraph to describe \textit{``network slicing''} and \textit{``edge computing''} as 5G architectures, with a short definition. GPT3.5 identifies centralized, distributed and cloud RAN as architectures. GPT4, however, identifies SA and NSA, and provides a description for each to discern the fundamental difference. LLaMA identifies SA and NSA, followed by a distinction between the two by describing the SA, thus indirectly implying that NSA relies on existing infrastructure. 

Question 3 pertains to applications of mid-band 5G. Here, Bard elucidates on enhanced Mobile BroadBand (eMBB), Fixed-Wireless Access (FWA) and enterprise use-cases including Augmented/Virtual Reality (AR/VR), automation, machine learning and video streaming. GPT3.5 begins by calling out the advantages related to speed and coverage, and mentions possible applications like video streaming/calling, gaming and Internet of Things (IoT) devices. GPT4 prioritizes possible applications in its reply (includes eMBB, FWA, smart cities, connected vehicles and IoT devices) and then mentions the advantages of mid-band 5G. LLaMA calls out the subjective nature of use-cases, and hence prioritizes the advantages, followed by brief exemplary use-cases like connected cars, cities and IoT devices. We observe that applications may be subjective and non-restrictive, hence it is important for the user to be provided with the reasons that would help identify types of applications that can be enabled with mid-band 5G. However, GPT4 and Bard have elaborate answers compared to the others for this question.

\color{black} Question 4 relates to ability of modems to support LTE and 5G along with support for network slices. Here, Bard gives the right answer, the underlying technology is pointed out incorrectly as network slicing (and its advantages is elaborately described in another). Further, Bard answers that it is possible to connect to both LTE and 5G at same time through configurations,  making the answer/justification incorrect. GPT3.5 provides a short confirmation, but incorrectly refers to it as dual connectivity. It is of interest to note that both Bard and GPT3.5 are incorrectly associating steering with dual connectivity. GPT4 not only provides the right answer, but also the correct justification while referring to 3GPP release and appropriate justification of 5G network slicing based on the application. LLaMA though agrees it is possible, also incorrectly associates steering to multi-connectivity. We hypothesize co-occurrence of LTE and 5G can result in spurious correlation to dual-connectivity, hence these LLMs may incorrectly answer these. \color{black}

\subsubsection{MOS and Fleiss' kappa}
In order to quantify the assessment of responses obtained from various models, we asked SMEs to assess the responses obtained and individually rate them on a scale of $1-5$, where a response of $1$ corresponds to low relevance and $5$ corresponds to high relevance (technical correctness and completeness of responses). All SMEs have 10 or more years of experience in the telecom domain and have been actively involved with development of telecom AI use cases over the past few years. The Mean Opinion Score (MOS) \cite{itu2017vocabulary} obtained from independent evaluation by five SMEs for the respective questions (Q1-Q4) is shown in Table \ref{tab:mos}. 

\begin{table}[h]
\centering
\begin{tabular}{|c|c|c|c|c|}
\hline
Question & GPT3.5 & GPT4 & Bard & LLaMA \\ \hline
Q1    & 2.8     & 4.4      & 4.2  & 2.2   \\ \hline
Q2    & 2.6     & 5        & 3.8  & 3     \\ \hline
Q3    & 2.8     & 4.8      & 4.2  & 2.2   \\ \hline
Q4    & 3.4     & 4.2      & 4.2  & 2.2   \\ \hline
\textbf{Avg}   & \textbf{2.9}     & \textbf{4.6}      & \textbf{4.1}  & \textbf{2.4}   \\ \hline
\end{tabular}
\caption{Mean Opinion Score for questions} \label{tab:mos}
\end{table}

Across the questions, we find that the MOS is highest for GPT4 responses consistently, while it is lowest for LLaMA. We also compute inter-rater agreement score using Fleiss' kappa \cite{fleiss1971measuring}, which is obtained as $0.316$, indicating fair agreement. The scores across the questions are consistent though. It may be noted here that Fleiss' kappa has been computed across 16 categories (4 questions for 4 models).

For product-related queries, we discuss findings from E2. Question 1 pertains to number and type of expansion slots in Cradlepoint's E300 router. From the product datasheet, there are 3 expansion slots - for MC400 modular modem expansion, USB2.0 Type A and MC20BT expansion. Bard and GPT3.5 do not correctly identify the  expansion slots, in fact Bard generates a recipe to access the expansion slots. GPT4 correctly identifies two of the expansion slots and gives a link to the specification sheet. LLaMA too incorrectly states that one slot type option is available but does not give details. 

\begin{figure}
    \centering
    \includegraphics[scale=0.4]{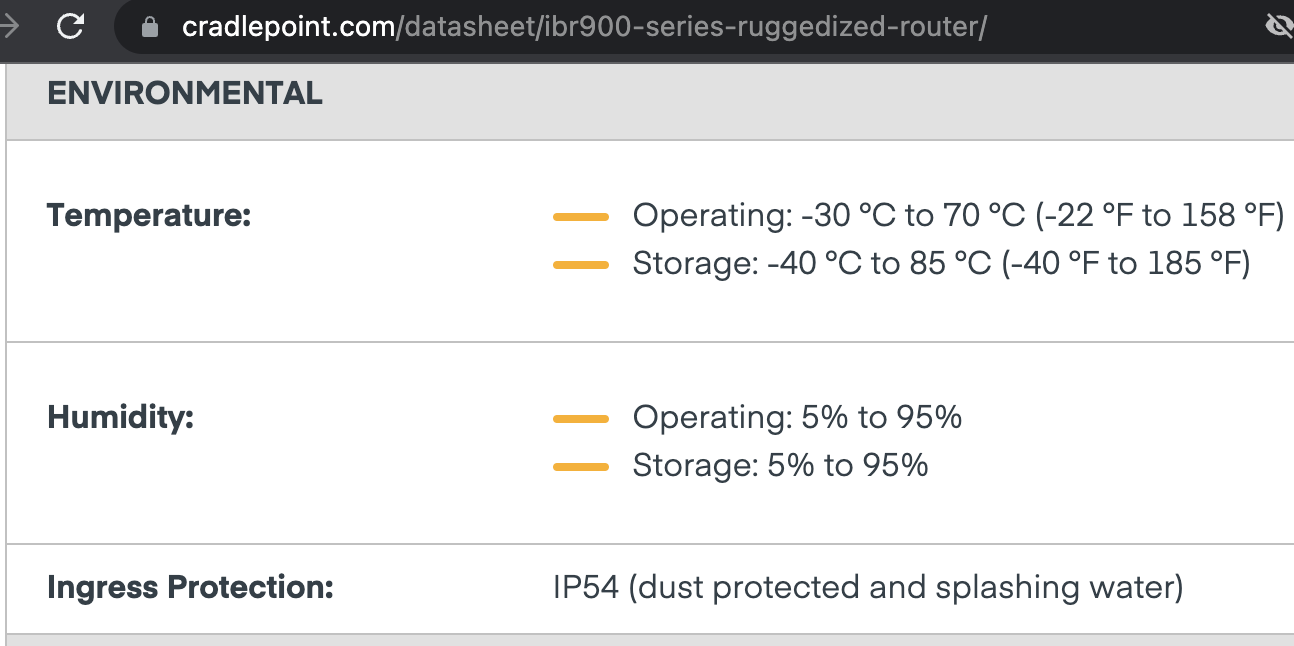}
    \caption{Ground truth for E2, Q2 [Source: \href{https://cradlepoint.com/datasheet/ibr900-series-ruggedized-router/}{IBR900 Datasheet}]}
    \label{fig:e2q2_gt}
\end{figure}

Question 2 pertains to operating conditions for Cradlepoint's IBR900 (a snapshot of product datasheet is shown in Figure \ref{fig:e2q2_gt}). It is possible that the content in product datasheets is updated, hence access to the latest product specifications datasheet is important for such queries to be answered correctly. However, the updates would pertain to adding certifications or regulatory approvals. It is very unlikely that operating temperatures would change. Hence, this specification attribute has been chosen to test for factual accuracy of the responses.

All the LLMs provide factually incorrect ranges. Bard provides temperature, humidity, altitude, shock and vibration but the ranges are incorrect. GPT3.5 also provides power input and consumption, while GPT4 provides operating and storage temperature and humidity. LLaMA provides generic temperature range and humidity rate from Cradlepoint. Q3-Q5 of E2 relate to recipe generation (RQ3) and is discussed in the corresponding sub-section \ref{ssec:rq3}. 

E2, Q6 pertains to finding products of a specific classification (ruggedized routers) from multiple products. E2, Q7 relates to comparison of product specifications for two routers (E300 and IBR900). Both of these relate to aggregation of information from multiple products. A correct response to this question must list 8 products. Responding to E2, Q6, Bard suggests IBR1700, IBR900 and R2100 (implying 100\% on accuracy, 37.5\% on completeness); GPT3.5 suggests IBR1700, IBR900, IBR600C and IBR200 (100\% accuracy, 50\% completeness); GPT4 suggests IBR900, IBR1700 and IBR600C (implying 100\% on accuracy, 37.5\% on completeness); while LLaMA indicates IBR3000/300B series and ECM managed routers, along with MC4000, NetCloud Essential and NetCloud Plus (implying 0\% accuracy, 0\% completeness). Although none of the LLMs list the complete set, access to latest information about products determines the response of the model, and hence the low numbers on completeness of the response.

With respect to Q7, Bard generates a table of 12 specifications and compares them across the products, and also includes a textual comparison of important features. 4 of the specification fields are not relevant to the router products considered and have no source in Cradlepoint domain. Of the remaining 8, none of the answers are correct. GPT3.5 compares using 6 specification fields as bullet lists, all of them are relevant. Out of these, only 2 answers are factually correct. GPT4 compares through 8 specification fields (all relevant and 4 answers being correct for both products) and provides links to specification sheets (although domain part of url is correct, there is no such sheet available). Surprisingly, LLaMA points to a non-existent link that is intended to be comparing the products. We also observe that a tabular representation of the specifications is useful for the reader for a comparison, such as the one provided by Bard.\\

\noindent\fcolorbox{black}{lightgray}{%
    \parbox{\linewidth}{%
        \textbf{Conclusions for RQ1:} GPT4 and Bard provide better responses for domain-specific questions. Hallucination is a potential risk for cases with high specificity, such as product models, names, specifications and components. Fine-tuning models with domain data corpus and additional pre and post-processing may alleviate some risks. Access to recent data is required for addressing completeness in the responses. 
    }%
}

\subsection{RQ2}
\label{ssec:rq2}
\begin{figure}[hbtp]
    \centering
    \begin{subfigure}[b]{\linewidth}
        \includegraphics[width=\linewidth]{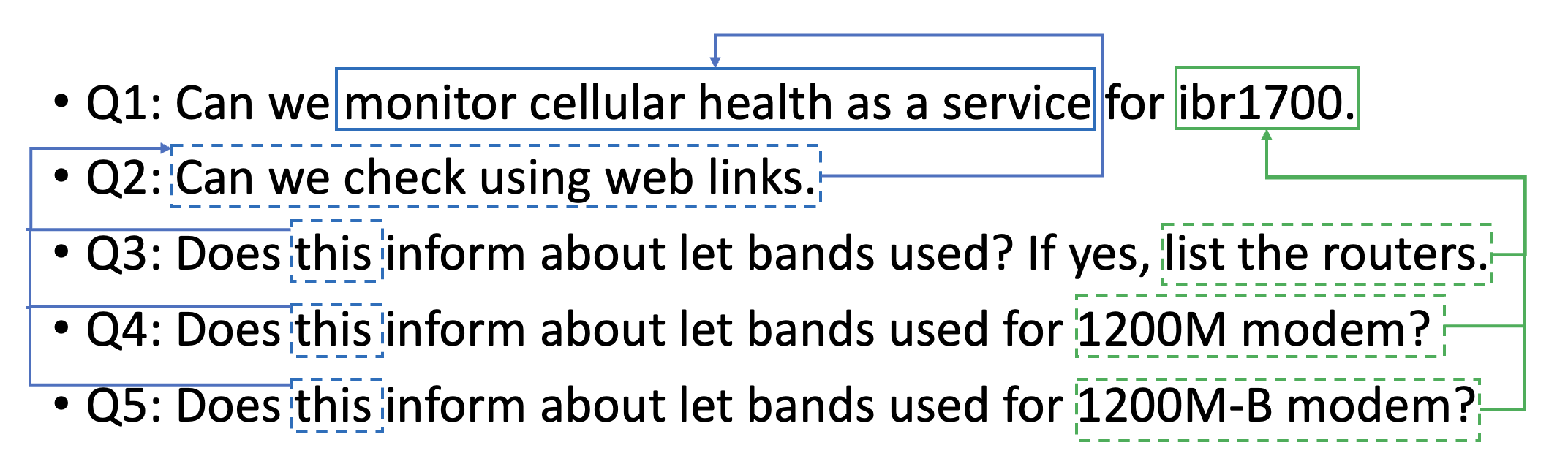}
        \caption{Context Continuity - E3}
        \label{fig:cc1}
    \end{subfigure}
    ~ 
    \begin{subfigure}[b]{\linewidth}
        \includegraphics[width=\linewidth]{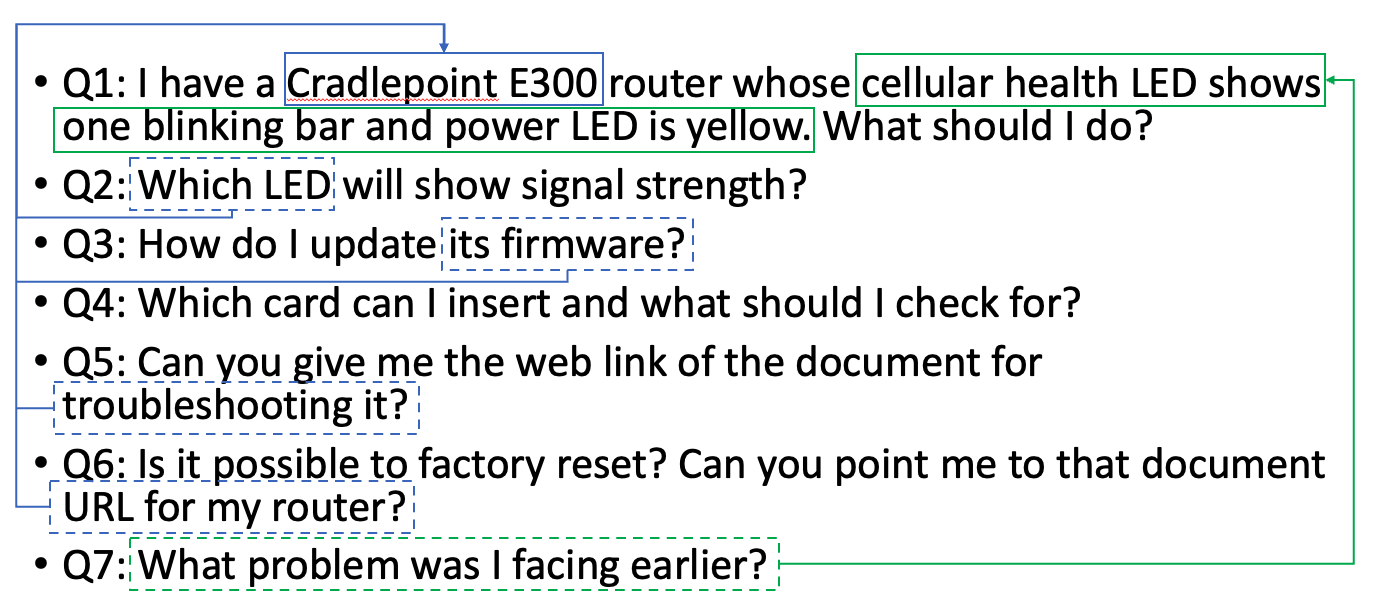}
        \caption{Context Continuity - E4}
        \label{fig:cc2}
    \end{subfigure}
    \caption{Context continuity - solid lines indicate context and dotted boxes connect the referring segments from queries.}\label{fig:cc}
\end{figure}

We address context retention aspects for conversations in this sub-section, that map to E3 and E4. The expected context for the set of queries is indicated in Figure \ref{fig:cc}. The boxes with solid lines indicate the context being referred to, and boxes with dotted lines indicate the referring segments from following queries. Each set of referring and referred contexts is indicated in a different color (blue, green). For our experiments, we consider retaining context across previous 2-3 questions as \textit{``short-term''} context retention, while references following more questions would be considered as \textit{``long-term''} context retention. We also point out here that the conversational interfaces used for our experiments do not provide for any explicit context-retention configuration.

In Fig. \ref{fig:cc1}, we refer to context of monitoring cellular health as a service in Q2, while other questions refer to context of router (product) and models. However, the word ``this'' here is ambiguous and can either refer to any Cradlepoint web link or the link provided in response to E3, Q2. This scenario is typically encountered in a human interaction and one would expect a counter-question for clarification. However, current LLMs do not ask for clarification and hence in this report, we evaluate only the LLM response inspite of the ambiguity. Fig. \ref{fig:cc2} illustrates a typical troubleshooting scenario, that can have both short (blue) and long (green)-term context references.

For E3, Bard is able to retain the context throughout the conversation, and all replies are relevant for IBR1700. It also provides the specific LTE bands supported (88\% accurate and 69\% complete). Bard doesn't generate any URLs. GPT3.5 and GPT4 also retain context, but they do not list the specific LTE bands that are supported. However, the links provided are incorrect (and do not exist).  Responding to E3, Q3-Q5 set, Bard indicates that LTE band information is available through the NetCloud account. GPT3.5 and GPT4 also indicate the same. This is incorrect. With
LLaMA, context retention is seen between E3, Q1-Q2. The links provided  do not exist and hence are incorrect. LLaMA does not retain the context between E3,Q2 to E3,Q3 and responds by listing three Cradlepoint routers which support LTE (66\% accurate) along with the countries supported. Of the two correct routers identified, one is currently discontinued.  We do not assess completeness as this requires access to latest product list. Subsequently, LLaMA responds to E3, Q4 and E3, Q5 keeping response of E3, Q3 as the context. Hence, technically, the answers are correct as it only addresses LTE bands supported.

For E4, Bard retains context related to the router (E300) for which the troubleshooting questions are posed. It points to relevant steps for firmware upgrade with reference to the router (post disambiguation), though the exact steps for SIM Card are not retrieved. It politely refuses to return a document link for troubleshooting the product (Q5). The factory reset steps (Q6) are not specific to the product and the response does not reference the product in Q1. For long term context (Q7), the top response from Bard does not reflect the original problem and the reply indicates that the steps suggested have \textit{``solved''} the problem, without any such indication being explicit in the query. Overall, 6 of 7 responses from Bard are categorized as acceptable by SME. 

GPT3.5 has a better response to the E4, Q1 by suggesting steps to debug power source. The context beyond E4, Q3 is generic, does not have product specific responses and returns generic router document links for Q6 and Q7. It however, does retain the long-term context, as it replies with the original problem as described by the user for response to Q7. The links provided do not exist even though the domain is correct. Overall, 3 of the 7 responses from GPT3.5 are categorized as acceptable. 

The limitations of GPT3.5 on context retention are overcome in the responses from GPT4, where responses and document links specific to the product are returned (though the link is incorrect and does not exist). The context is also retained through Q7, where the response disambiguates the two distinct issues that were described by the user in Q1. 6 of the 7 responses from GPT4 for E4 are SME acceptable. 

LLaMA's responses for E4,Q1 seems largely deviant to the user query. It also does not explicitly retain product-specific context for Q1 and Q2, but it does so for Q3 (the answer to Q3 itself may be incorrect). For Q4, it asks for more context and does not give any reply (which is a desired feature), while for Q5 it returns irrelevant URLs. For Q6, it appears it has lost context to disambiguate the router and hence, it returns links for other products, while for Q7 the issue is incorrectly re-stated with additional statements, and it claims to have raised a support ticket for a technician to attend to the issue (while also providing a reference number for the same). 3 of 7 responses from LLaMA are categorized as acceptable by SME. Based on these, we primarily find GPT4 and Bard's responses for context retention to be more persistent and relevant.


\noindent\fcolorbox{black}{lightgray}{%
    \parbox{\linewidth}{%
    \textbf{Conclusions for RQ2:} Amongst the LLMs considered in this work, GPT4 provide both long and short term context retention and can potentially be useful in scenarios which require troubleshooting. However, when it comes to context retention, it is worth noting that ambiguous queries can result in ambiguous answers which is true philosophically also, even with humans. Hence, context disambiguation must be evaluated as a shared responsibility between the user and LLM. In addition, while the current technology push is towards answering all the questions (as "interpreted" by the LLM), it is desirable that the LLM asks the user for clarification (like LLaMA) or politely refuse to provide any URL (like Bard). 
    }%
}

\subsection{RQ3}
\label{ssec:rq3}
We discuss results for queries that expect recipe generation from the models, such as E2, Q3. The response from Bard provides 6 steps, including configuring the router through the web-based interface. We find this is reliable, and conforms to the process of router setup for Cradlepoint products (except that there is no mention of SIM card for cellular connectivity; this, however, may not be the primary concern).  GPT3.5 provides 6 steps, and instructs on activation of router in steps 3 \& 4, which are not adherent to Cradlepoint process. It provides a link to quick start guide where domain is correct, but actual link doesn't exist. GPT4 provides 10 steps for setting up IBR900. The steps till hardware setup are accurate, but is also incorrect regarding the access/activation step. The steps provided by GPT4 are more detailed and specific in comparison to GPT3.5. The URL link provided by GPT4 is also incorrect, but the domain is correct. LLaMA provides a URL, which is ``safer'' but the URL is non-existent. For a quick comparison of LLMs within this experiment, the accuracy of the steps is 100\%, 50\%, 70\%, 0\% for Bard, GPT3.5, GPT4 and LLaMA respectively. 

For evaluating hallucination, we refer to responses for E2, Q4 and Q5. GPT4 too identifies the incorrect product and offers information on another product (IBR600C) as an alternative. These product identification issues may also possibly be indicative of the significance of character-level representations in such models \cite[pp 7, col 1, para 2]{moradi2021evaluating}. GPT3.5 and Bard are, however, unable to distinguish the incorrect product name and generate responses without any indication of their factual incorrectness. LLaMA is unable to identify the fictitious product and indicates that it does not have information on that product as its training cut-off date was September 2021. In this case, we find that GPT4 and LLaMA are comparatively reliable (less hallucination) with their responses.


\noindent\fcolorbox{black}{lightgray}{%
    \parbox{\linewidth}{%
    \textbf{Conclusions for RQ3:} We can view two major aspects for reliability. First, the LLM steps for recipe generation may not be completely accurate for any LLM, even if there are guardrails provided in the prompt to prevent hallucination. This could be due to the popularity bias from competitor products and their approach for any task. It is desirable that LLM outputs are more reliable (like that of Bard), or LLM redirects the users to the source URL (like in LLaMA), with the URL being a valid one. Second, the LLM may not be able to discern products if its architecture doesn't include character-level representation and are mostly based on tokens. It is possible that due to these shortcomings, LLMs may not be directly useful in scenarios which require reliable answers. 
    }%
}

\subsection{RQ4}

These set of experiments (E3, Q4-Q5 and E5) aim to evaluate robustness of models to input perturbations introduced through deliberate spelling and language typos, the typos covering both English language and those pertaining to telecom domain terminology and products. The errors in E3 (Q4 and Q5) are resolved unambiguously across all the models evaluated. Bard is able to resolve all errors in the questions in E5. GPT3.5 fails on correcting the errors in E5, Q1 and interacts with the user to clarify the query. GPT4 resolves the error, but is not confident and hence asks for a clarification from the user. Both GPT3.5 and GPT4 resolve errors in the rest of the questions. LLaMA resolves spell errors for Q1 and Q4 only. However, response to these are incorrect as  they do not relate to Cradlepoint's products (despite having the prompt as a guardrail). The responses for Windows machines (Q1) and generic Digital Subscriber Line (DSL) modems and other products has partial similarity in product names with Cradlepoint products. In Q5, LLaMA responds saying ``IKE'' does not relate to networking. In the subsequent question (Q6), it is able to correctly resolve IKE abbreviation, it does not provide any configuration steps (maybe because it could not resolve ``configure'' perturbation). 

\vspace{3mm}

\noindent\fcolorbox{black}{lightgray}{%
    \parbox{\linewidth}{%
        \textbf{Conclusions for RQ4}: From the limited experiments, we observe that Bard is more robust to input perturbations. GPT4 is also able to correct the typos and also has the desirable characteristic of clarifying from the user when it is not confident (hypothesis). LLaMA seems to be sensitive to input perturbations. It is also observed that the guard rails introduced through prompts do not hold for LLaMA when the query has input perturbations. 
    }%
}
\section{Conclusions and Future Work \label{sec:conc}}
We evaluated four popular LLMs for typical chatbot requirements in the enterprise wireless domain, utilizing Cradlepoint offerings for experiments. We used limited datasets so as to understand the suitability of LLMs, before testing it out on larger datasets. The learnings presented in our work would be useful for other domains as well. We have also introduced prompts that serve as guardrails to minimize hallucination. 

Some noteworthy observations are as follows: 
\begin{itemize}
    \item It is observed that for domain related questions, Bard and GPT4 show promise with respect to accuracy and could be useful. However, it is inevitable to lean towards fine-tuning for performance improvements. 
    \item GPT4 is found to be most suitable for both short and long-term context retention, based on our assessment. A detailed study in this direction is desirable.
    \item It is desirable that LLMs ask clarifications when the user query is ambiguous. Of course, the caveat to determine if the query is ambiguous is another research problem. 
    \item It is desirable that LLMs refuse to provide URLs because their operational mode is restricted to training/inference. It is note-worthy that GPT3.5 and GPT4 can include the domain specified in the prompt in their responses, though exact URL is incorrect. We hypothesize that plugins (or other interface approaches) could facilitate this.
    \item Summarization abilities require reliability. Bard seems to be more closer to the requirements assessed here. A detailed study is required to further ascertain these initial observations. 
    \item It is desirable that LLMs do not hallucinate product name(s) in such use cases. This may requires LLMs to have character aware features to discern such aspects.
    \item LLMs appear to be quite robust to domain related spelling perturbations. This could be because of the context obtained from the query or previous conversation(s). 
\end{itemize}
Based on our findings, we are inclined to conclude that such LLMs can't be used directly (in the form made available publicly) for enterprise use-cases. It is imperative that these models are fine-tuned for domain specific tasks. Also, it has been hypothesized that spurious correlations could be the cause of some wrong answers, such as those seen in association of 4G \& 5G with dual connectivity in all LLMs except GPT4. Detailed experiments need to be conducted to ascertain and mitigate these risks. Licensing, data security and privacy aspects, deployment costs and associated constraints are practical considerations that also need be assessed and evaluated for enterprise-grade applications. 

\section*{Acknowledgments}
We thank Anubhav Arora and Natarajan Venkataraman from Cradlepoint for their timely and valuable inputs. We are grateful to Venkatesan Pradeep, Ram Kumar Sharma, Pushpendra Sharma, Satish Kumar Kolli and K. N. S. S. Yoganand for their evaluation and inputs on the domain specific responses. We thank Ramana Murthy, Chandramouli Sargor and Thirumaran Ekambaram for facilitating this evaluation study and for their unwavering support.
\bibliographystyle{unsrt}  
\bibliography{ref}

\end{document}